\documentclass[preprint,12pt]{elsarticle}
\usepackage{amssymb}
\usepackage{amsmath}

\journal{Physics Letters A}

\begin{document}

\begin{frontmatter}

\title{Deconvolution of 3-D Gaussian kernels}

\author{Z.~K.~Silagadze}

\address{ Budker Institute of Nuclear Physics and Novosibirsk State 
University, 630 090, Novosibirsk, Russia.}

\begin{abstract}
Ulmer and Kaissl formulas for the deconvolution of one-dimensional Gaussian 
kernels are generalized to the three-dimensional case. The generalization 
is based on the use of the scalar version of the Grad's multivariate Hermite 
polynomials which can be expressed through ordinary Hermite polynomials.
\end{abstract}


\end{frontmatter}

\section{Introduction}
Mathematical problem of deconvolution of Gaussian kernels was considered
in \citep{1}. In one-dimensional case, such a deconvolution (inverse problem)
is equivalent to the solution of the  Fredholm-type integral equation
\begin{equation}
\phi(x)=\int\limits_{-\infty}^\infty K(x-\xi;\sigma) \,\rho(\xi)\, d\xi,
\label{eq1}
\end{equation}
where $\phi(x)$ is the known function, typically a detector response in 
a some measurement process of the unknown signal $\rho(\xi)$ which is blurred 
by the finite resolution $\sigma$ of the detector. This blurring is described
by the Gaussian kernel
\begin{equation}
K(x-\xi;\sigma)=\frac{1}{\sigma\sqrt{\pi}}\,e^{-\frac{(x-\xi)^2}{\sigma^2}}.
\label{eq2}
\end{equation} 
The integral transform (Gaussian convolution) (\ref{eq1}) arises in various 
problems of applied physics, for example in calculations of transverse 
profiles of photon, proton or electron beams in medical applications
\citep{1,2,3,4}.

Usually the integral equation (\ref{eq1}) is solved  by  a Fourier transform
method. It is well known that, due to a presence of a fast growing Gaussian 
function in the deconvolution integral in this case, this method may imply 
ill-posed inverse problems \citep{1}. Therefore alternative methods of
deconvolution of Gaussian kernels, avoiding ill-posed inverse problems,
were developed \citep{1,2}.

In particular, Ulmer and Kaissl provide in \citep{1} two different expressions
for the inverse kernel $K^{-1}(x-\xi;\sigma)$ defined through the relation
\begin{equation}
\rho(x)=\int\limits_{-\infty}^\infty K^{-1}(x-\xi;\sigma) \,\phi(\xi)\, d\xi.
\label{eq3}
\end{equation}
In the next section we reproduce their results by a method which can be 
easily generalized to the three-dimensional case.

\section{Deconvolution of 1-D Gaussian kernels}
Our starting point will be an identity \citep{5}
\begin{equation}
K(x;\sigma)=e^{\frac{\sigma^2}{4}\,\frac{d^2}{dx^2}}\,\delta(x),
\label{eq4}
\end{equation}
which can be easily proved by using the Fourier integral representation of the 
Dirac's delta function:
\begin{equation}
\delta(x)=\frac{1}{2\pi}\int\limits_{-\infty}^\infty e^{ipx}\,dp.
\label{eq5}
\end{equation}
Because of (\ref{eq4}), we can write
\begin{eqnarray} &&
\phi(x)=\int\limits_{-\infty}^\infty \rho(\xi)\, e^{\frac{\sigma^2}{4}\,
\frac{d^2}{d\xi^2}}\,\delta(x-\xi)\,d\xi= \nonumber \\ &&
\int\limits_{-\infty}^\infty 
\delta(x-\xi)\,e^{\frac{\sigma^2}{4}\,\frac{d^2}{d\xi^2}}\,\rho(\xi)\,d\xi=
e^{\frac{\sigma^2}{4}\,\frac{d^2}{dx^2}}\,\rho(x). 
\label{eq6}
\end{eqnarray}
Therefore the integral transform (\ref{eq1}) can be written in the operator
form
\begin{equation}
\phi(x)=\hat K \rho(x),\;\;\; \hat K=e^{\frac{\sigma^2}{4}\,\frac{d^2}{dx^2}}.
\label{eq7}
\end{equation}
Hence its inverse is given by
\begin{equation}
\rho(x)=\hat K^{-1} \phi(x),\;\;\; \hat K^{-1}=e^{-\frac{\sigma^2}{4}\,
\frac{d^2}{dx^2}}.
\label{eq8}
\end{equation}
Now we apply the same trick as in (\ref{eq6}):
$$\rho(x)=\int\limits_{-\infty}^\infty \delta(x-\xi)\,e^{-\frac{\sigma^2}{4}\,
\frac{d^2}{d\xi^2}}\phi(\xi)=\int\limits_{-\infty}^\infty \phi(\xi)\,
e^{-\frac{\sigma^2}{4}\,\frac{d^2}{d\xi^2}}\delta(x-\xi),$$
which is the same as
\begin{equation}
\rho(x)=\int\limits_{-\infty}^\infty \phi(\xi)\,e^{-\frac{\sigma^2}{4}\,
\frac{d^2}{dx^2}}\delta(x-\xi).
\label{eq9}
\end{equation}
but, because of (\ref{eq4}), we have
\begin{equation}
e^{-\frac{\sigma^2}{4}\,\frac{d^2}{dx^2}}\delta(x-\xi)=
e^{-\frac{\sigma^2}{2}\,\frac{d^2}{dx^2}}\,
e^{\frac{\sigma^2}{4}\,\frac{d^2}{dx^2}}\delta(x-\xi)=
e^{-\frac{\sigma^2}{2}\,\frac{d^2}{dx^2}}K(x-\xi; \sigma).
\label{eq10}
\end{equation}
Therefore 
$$\rho(x)=\int\limits_{-\infty}^\infty \phi(\xi)\,
e^{-\frac{\sigma^2}{2}\,\frac{d^2}{dx^2}}K(x-\xi; \sigma),$$
which shows that
\begin{equation}
K^{-1}(x-\xi; \sigma)=e^{-\frac{\sigma^2}{2}\,\frac{d^2}{dx^2}}K(x-\xi; 
\sigma)=\sum\limits_{n=0}^\infty \frac{(-\sigma^2)^n}{2^n\,n!}\,
\frac{d^{2n}}{dx^{2n}}K(x-\xi; \sigma).
\label{eq11}
\end{equation}
Thanks to the Rodrigues formula for Hermite polynomials \citep{6}
\begin{equation}
H_n(x)=(-1)^n\,e^{x^2}\,\frac{d^n}{dx^n}e^{-x^2},
\label{eq12}
\end{equation}
derivatives of the Gaussian function in (\ref{eq11}) can be expressed through
Hermite polynomials:
\begin{equation}
\frac{d^{2n}}{dx^{2n}}K(x-\xi; \sigma)=\sigma^{-2n}\,H_{2n}\left (
\frac{x-\xi}{\sigma}\right )\,K(x-\xi; \sigma),
\label{eq13}
\end{equation}
and we finally obtain
\begin{equation}
K^{-1}(x-\xi; \sigma)=\sum\limits_{n=0}^\infty 
\frac{(-1)^n}{2^n\,n!}\;H_{2n}\left ( \frac{x-\xi}{\sigma}\right )\,
K(x-\xi; \sigma).
\label{eq14}
\end{equation}
Another version of the inverse Gaussian kernel can be obtained by the use
of the following identity
\begin{equation}
e^{-\frac{\sigma^2}{4}\,\frac{d^2}{dx^2}}\delta(x-\xi)=
\delta(x-\xi)+\left (e^{-\frac{\sigma^2}{2}\,\frac{d^2}{dx^2}}
-e^{-\frac{\sigma^2}{4}\,\frac{d^2}{dx^2}}\right )
e^{\frac{\sigma^2}{4}\,\frac{d^2}{dx^2}}\,\delta(x-\xi).
\label{eq15}
\end{equation}
Using this identity and the identity (\ref{eq4}) in (\ref{eq9}), we get
\begin{eqnarray} &&
\rho(x)=\int\limits_{-\infty}^\infty \phi(\xi)\left [ \delta(x-\xi)+
\left (e^{-\frac{\sigma^2}{2}\,\frac{d^2}{dx^2}}
-e^{-\frac{\sigma^2}{4}\,\frac{d^2}{dx^2}}\right )K(x-\xi;\sigma)\right]
\, d\xi= \nonumber \\ &&
\int\limits_{-\infty}^\infty \phi(\xi)\left [ \delta(x-\xi)+
\sum\limits_{n=1}^\infty 
\frac{(-\sigma^2)^n}{4^n\,n!}\,(2^n-1)\,\frac{d^{2n}}{dx^{2n}}K(x-\xi; \sigma)
\right ].
\label{eq16}
\end{eqnarray}
Therefore, in light of (\ref{eq13}), we have
\begin{equation}
K^{-1}(x-\xi; \sigma)=\delta(x-\xi)\,+\,\sum\limits_{n=1}^\infty
\frac{(-1)^n}{4^n\,n!}\,(2^n-1)\;H_{2n}\left ( \frac{x-\xi}{\sigma}\right )\,
K(x-\xi; \sigma).
\label{eq17}
\end{equation}
Equations (\ref{eq14}) and (\ref{eq17}) were first obtained (with some typos) 
in \citep{1}.

\section{Deconvolution of 3-D Gaussian kernels}
Generalization to the three-dimensional case is now straightforward. All what 
is needed is to change the differential operator $\frac{d^2}{dx^2}$ by its 
three-dimensio\-nal version (Laplacian) $\Delta=\boldsymbol{\nabla}^2$, change
$x$ and $\xi$ by the corresponding three-dimensional vectors $\mathbf{r}$
and  $\boldsymbol{\xi}$, and use a three-dimensional Gaussian kernel
\begin{equation}
K(\mathbf{r}-\boldsymbol{\xi};\sigma)=\frac{1}{\sigma^3\sqrt{\pi^3}}\,
e^{-\frac{(\mathbf{r}-\boldsymbol{\xi})^2}{\sigma^2}}.
\label{eq18}
\end{equation}
Then, instead of (\ref{eq11}), we will end up with the equation
\begin{equation}
K^{-1}(\mathbf{r}-\boldsymbol{\xi};\sigma)=e^{-\frac{\sigma^2}{2}\,
\Delta}K(\mathbf{r}-\boldsymbol{\xi};\sigma)\, = \,
\sum\limits_{n=0}^\infty 
\frac{(-\sigma^2)^n}{2^n\,n!}\,
\Delta^n K(\mathbf{r}-\boldsymbol{\xi};\sigma).
\label{eq19}
\end{equation}
However now the action of the powers of Laplacian $\Delta=\boldsymbol
{\nabla}^2$ on the three-dimensional Gaussian function cannot be as simply 
expressed in terms of the Hermite polynomials, as in the one-dimensional
case. In \citep{2} the following approach was suggested. Using
$$K(\mathbf{r}-\boldsymbol{\xi};\sigma)=K(x-\xi_x; \sigma)\,
K(y-\xi_y; \sigma)\,K(z-\xi_z; \sigma),$$
and
\begin{equation}
e^{-\frac{\sigma^2}{2}\,\Delta}=e^{-\frac{\sigma^2}{2}\,\frac{d^2}{dx^2}}\;
e^{-\frac{\sigma^2}{2}\,\frac{d^2}{dy^2}}\;
e^{-\frac{\sigma^2}{2}\,\frac{d^2}{dz^2}},
\label{eq19A}
\end{equation}
we get from the first equation of (\ref{eq19})
\begin{equation}
K^{-1}(\mathbf{r}-\boldsymbol{\xi};\sigma)=F_1F_2F_3
\,K(\mathbf{r}-\boldsymbol{\xi};\sigma),
\label{eq20}
\end{equation}
where
\begin{eqnarray} &&
F_1=\sum\limits_{n_1=0}^\infty 
\frac{(-1)^{n_1}}{2^{n_1}\,n_1!}\;H_{2n_1}\left ( \frac{x-\xi_x}{\sigma}
\right ),\;F_2=\sum\limits_{n_2=0}^\infty \frac{(-1)^{n_2}}{2^{n_2}\,n_2!}\;
H_{2n_2}\left ( \frac{y-\xi_y}{\sigma}\right ),\nonumber \\ && 
F_3=\sum\limits_{n_3=0}^
\infty \frac{(-1)^{n_3}}{2^{n_3}\,n_3!}\;H_{2n_3}\left ( \frac{z-\xi_z}
{\sigma}\right ).
\label{eq21}
\end{eqnarray}
Yet this form of inverse kernel doesn't seem convenient in applications.
Especially if we note that a more straightforward generalization of
equations (\ref{eq14}) and (\ref{eq17}) is possible if we use multivariate
and tensorial generalization of Hermite polynomials introduced by Grad
\citep{7,8}. Grad's definition generalizes the Rodrigues formula for the 
univariate Hermite polynomials (\ref{eq12}):
\begin{equation}
{\cal H}^{(n)}_{i_1i_2\ldots i_n}(\mathbf{r})=(-1)^n\,e^{\mathbf{r}^2}\,
\boldsymbol{\nabla}_{i_1}\boldsymbol{\nabla}_{i_2}\cdots\boldsymbol{\nabla}_
{i_n}e^{-\mathbf{r}^2}.
\label{eq22}
\end{equation} 
Using this definition, we easily get straightforward generalization of the
equation (\ref{eq14}) in the form
\begin{equation}
K^{-1}(\mathbf{r}-\boldsymbol{\xi};\sigma)=\sum\limits_{n=0}^\infty 
\frac{(-1)^n}{2^n\,n!}\;{\cal H}_{2n}\left ( \frac{(\mathbf{r}-
\boldsymbol{\xi})^2}{\sigma^2}\right )\,K(\mathbf{r}-\boldsymbol{\xi};\sigma),
\label{eq23}
\end{equation}
where 
\begin{equation}
{\cal H}_{2n}\left (\frac{(\mathbf{r}-\boldsymbol{\xi})^2}{\sigma^2}\right )=
\delta_{i_1i_2}\delta_{i_3i_4}\cdots \delta_{i_{2n-1}i_{2n}}{\cal H}^{(2n)}_
{i_1 i_2\ldots i_{2n-1}i_{2n}}(\mathbf{r}-\boldsymbol{\xi};\sigma)
\label{eq24}
\end{equation}
is the completely contracted version of the multivariate Hermite polynomials
(the so called scalar, or irreducible, Hermite polynomials \citep{9}).

Analogously, the generalization of the equation (\ref{eq17}) looks like
\begin{equation}
K^{-1}(\mathbf{r}-\boldsymbol{\xi}; \sigma)=\delta(\mathbf{r}-
\boldsymbol{\xi})\,+\,\sum\limits_{n=1}^\infty
\frac{(-1)^n}{4^n\,n!}\,(2^n-1)\;{\cal H}_{2n}\left ( \frac{(\mathbf{r}-
\boldsymbol{\xi})^2}{\sigma^2}\right )\,
K(\mathbf{r}-\boldsymbol{\xi};\sigma).
\label{eq25}
\end{equation}

\section{Calculation of scalar Hermite polynomials}
Scalar Hermite polynomials can be expressed in terms of the classical 
(associated) Laguerre polynomials \citep{9,10}. Below we elaborate this
connection and show that, in fact, Scalar Hermite polynomials can be 
expressed through the ordinary Hermite polynomials.

Let us introduce the following operators \citep{11}
\begin{equation}
\hat E=\frac{1}{2}\,r^2,\;\;\hat F=-\frac{1}{2}\,\Delta,\;\;
\hat H=\mathbf{r}\cdot\boldsymbol{\nabla}+\frac{3}{2}.
\label{eq26}
\end{equation}
Using the commutator
$$[\nabla_i,\,r]=\frac{r_i}{r},$$
it can be easily checked that these operators obey commutation relations
of the $\mathfrak{sl}(2,\mathbb C)$ Lie algebra \citep{11}:
\begin{equation}
[\hat H,\,\hat E]=2\hat E,\;\;\;[\hat H,\,\hat F]=-2\hat F,\;\;\;
[\hat E,\,\hat F]=\hat H.
\label{eq27}
\end{equation}
It is clear from (\ref{eq22}) and (\ref{eq24}) that
\begin{equation}
{\cal H}_{2n}(\mathbf{r}^2)=e^{r^2}\,\Delta^n\,e^{-r^2}=
e^{\hat{r}^2}\,\Delta^n\,e^{-\hat{r}^2}(1)=\left (e^{\hat{r}^2}\,\Delta \,
e^{-\hat{r}^2}\right )^n(1),
\label{eq28}
\end{equation}
where $\hat r$ is just a multiplication operator: $\hat r g(r)=rg(r)$ for
any function $g(r)$, and an expression $\hat A (1)$ means that the operator
$\hat A$ acts on the function $g(r)=1$. The trick used here is borrowed from
\citep{12} and it is useful because the last expression in (\ref{eq28})
allows to use the operator identity (Baker-Hausdorff formula)  \citep{13}
\begin{equation}
e^{\hat A}\,\hat B\,e^{-\hat A}=\hat B+[\hat A,\,\hat B]+
\frac{1}{2!}[\hat A,[\hat A,\,\hat B]]+\frac{1}{3!}[\hat A,[\hat A,[\hat A\,
\hat B]]]+\cdots
\label{eq29}
\end{equation}
With the help of the commutation relations (\ref{eq27}), we can easily check
that in the case of the last expression in (\ref{eq28}) the corresponding 
Baker-Hausdorff series (\ref{eq29}) in fact terminates and we end up with 
\begin{equation}
e^{\hat{r}^2}\,\Delta \,e^{-\hat{r}^2}=e^{2\hat E}(-2\hat F)e^{-2\hat E}=
-2\hat F-4\hat H +8\hat E.
\label{eq30}
\end{equation} 
Analogously, we have
\begin{equation}
e^{-\frac{\Delta}{4}}\,(4r^2)^n=e^{-\frac{\Delta}{4}}\,(4\hat{r}^2)^n\,
e^{\frac{\Delta}{4}}(1)=\left (e^{-\frac{\Delta}{4}}\,(4\hat{r}^2)\,
e^{\frac{\Delta}{4}}\right )^n(1),
\label{eq31}
\end{equation}
and
\begin{equation}
e^{-\frac{\Delta}{4}}\,(4\hat{r}^2)\,e^{\frac{\Delta}{4}}=e^{\frac{1}{2}\hat F}
\,(8\hat E)\,e^{-\frac{1}{2}\hat F}=8\hat E-4\hat H-2\hat F.
\label{eq32}
\end{equation}
Comparing (\ref{eq32}) and (\ref{eq30}), we see from (\ref{eq28}) and 
(\ref{eq31}) that
\begin{equation}
{\cal H}_{2n}(\mathbf{r}^2)=e^{-\frac{\Delta}{4}}\,(4r^2)^n=
e^{-\frac{\Delta}{4}}\,(2r)^{2n}.
\label{eq33}
\end{equation}
Note that the relation (\ref{eq33}) is completely analogous to an alternative
expression for the usual Hermite polynomials \citep{14}
\begin{equation}
H_n(x)=e^{-\frac{1}{4}\,\frac{d}{dx}}\,(2x)^n.
\label{eq34}
\end{equation}
Although in some calculations (\ref{eq34}) is more useful than the Rodrigues 
formula, it was little known for a long time (see, for example, 
\citep{15,16}). 

Using
$$\Delta\,r^n=\frac{1}{r^2}\,\frac{d}{dr}\left(r^2\,\frac{dr^n}{dr}\right)=
n(n+1)\,r^{n-2},$$
it can be proved by induction that
\begin{equation}
\Delta^m\,r^{2n}=\frac{(2n+1)!}{(2n-2m+1)!}\,r^{2(n-m)},\;\;\mathrm{if}
\;\; n\ge m,
\label{eq35}
\end{equation}
and $\Delta^m\,r^{2n}=0$, if $m>n$. Then (\ref{eq33}) gives
$${\cal H}_{2n}(\mathbf{r}^2)=\sum\limits_{m=0}^\infty \frac{(-1)^m 4^{n-m}}
{m!}\,\Delta^m r^{2n}=\sum\limits_{m=0}^n \frac{(-1)^m 4^{n-m}}{m!}
\frac{(2n+1)!}{(2n-2m+1)!}\, r^{2(n-m)},$$
which, after introduction of a new summation index $i=n-m$, can be rewritten
as follows:
\begin{equation}
{\cal H}_{2n}(\mathbf{r}^2)=(-1)^n\sum\limits_{i=0}^n \frac{(-1)^i4^i}{(n-i)!}
\,\frac{(2n+1)!}{(2i+1)!}\,r^{2i}.
\label{eq36}
\end{equation}
In particular, we get for the first few scalar Hermite polynomials:
\begin{eqnarray} &&
{\cal H}_2(\mathbf{r}^2)=2\,(2r^2-3),\nonumber \\ &&
{\cal H}_4(\mathbf{r}^2)=4\,(4r^4-20r^2+15),\nonumber \\ &&
{\cal H}_6(\mathbf{r}^2)=8\,(8r^6-84r^4+210r^2-105),\nonumber \\ &&
{\cal H}_8(\mathbf{r}^2)=16\,(16r^8-288r^6+1512r^4-2520r^2+945).
\label{eq37}
\end{eqnarray}
In light of the Legendre duplication formula \citep{17}
\begin{equation}
(2z+1)!=2^{2z+1}\pi^{-1/2}z!\left (z+\frac{1}{2}\right )!,
\label{eq38}
\end{equation}
we have
$$\frac{4^i\,(2n+1)!}{(n-i)!\,(2i+1)!}=\frac{2^{2n}n!}{i!}\,
\frac{\left(n+\frac{1}{2}\right )!}{(n-i)!\left(i+\frac{1}{2}\right)!}=
\binom{n+\frac{1}{2}}{n-i}\,\frac{2^{2n}n!}{i!},$$
and (\ref{eq36}) takes the form
\begin{equation}
{\cal H}_{2n}(\mathbf{r}^2)=(-1)^n\,2^{2n}\,n!\sum\limits_{i=0}^n\frac{(-1)^i}
{i!}\,\binom{n+\frac{1}{2}}{n-i}\,(\mathbf{r}^2)^i.
\label{eq39}
\end{equation}
On the other hand, the associated Laguerre polynomials $L^k_n$ satisfy 
\citep{14}
\begin{equation}  
 L^k_n(x)=\sum\limits_{i=0}^n\frac{(-1)^i}{i!}\,\binom{n+k}{n-i}\,x^i.
\label{eq40}
\end{equation}
Comparing (\ref{eq39}) and (\ref{eq40}), we see that
\begin{equation}
{\cal H}_{2n}(\mathbf{r}^2)=(-1)^n\,2^{2n}\,n!\,L_n^{1/2}(\mathbf{r}^2).
\label{eq41}
\end{equation}
However \citep{14}
\begin{equation}
L_n^{1/2}(r^2)=\frac{(-1)^n}{2^{2n+1}\,n!\,r}\,H_{2n+1}(r),
\label{eq42}
\end{equation}
and we finally get a simple expression for the scalar Hermite polynomials
in terms of ordinary Hermite polynomials:
\begin{equation}
{\cal H}_{2n}(\mathbf{r}^2)=\frac{H_{2n+1}(r)}{2r}.
\label{eq43}
\end{equation}

\section{Concluding remarks}
As we see, the use of Grad's multivariate Hermite polynomials allows to 
perform a straightforward generalization of Ulmer and Kaissl formulas for the
inverse problem of a Gaussian convolution to the three-dimensional case.
In this generalization, the scalar Hermite polynomials play the same role
as ordinary Hermite polynomials in the Ulmer and Kaissl's one-dimensional 
formulas. The scalar Hermite polynomials by themselves can be expressed 
in terms of associated Laguerre polynomials and through them, rather
surprisingly, through ordinary Hermite polynomials. The latter connection
becomes less mysterious thanks to the new generating functions for even- 
and odd-Hermite polynomials \citep{18}:  
\begin{eqnarray} &&
F(x,t)=\sum\limits_{n=0}^\infty\frac{t^n}{n!}\,H_{2n}(x)=
\frac{1}{\sqrt{1+4t}}\,\exp{\left (\frac{4tx^2}{1+4t}\right )},\nonumber \\ &&
G(x,t)=\sum\limits_{n=0}^\infty\frac{t^n}{n!}\,H_{2n+1}(x)=
\frac{2x}{\sqrt{(1+4t)^3}}\,\exp{\left (\frac{4tx^2}{1+4t}\right )}.
\label{eq44}
\end{eqnarray}
Indeed, since
$$(2r)^{2n}=4^n(x^2+y^2+z^2)^n=\sum\limits_{n_1+n_2+n_3=n}
\frac{n!}{n_1!\,n_2!\,n_3!}\,(2x)^{2n_1}\,(2y)^{2n_1}\,(2z)^{2n_1},$$
(\ref{eq19A}), (\ref{eq33}) and (\ref{eq34}) imply that (\ref{eq43}) is 
equivalent to the curious identity
\begin{equation}
\sum\limits_{n_1+n_2+n_3=n}\frac{n!}{n_1!\,n_2!\,n_3!}\,H_{2n_1}(x)\,
H_{2n_2}(y)\,H_{2n_3}(z)=\frac{H_{2n+1}(r)}{2r},
\label{eq45}
\end{equation}
which, however, is a simple consequence of the generating functions 
(\ref{eq44}), because 
$$F(x,t)\,F(y,t)\,F(z,t)=\frac{G(r,t)}{2r}.$$

The task of deconvolution arises in many fields, such as geophysics, signal 
and image processing, underwater acoustics, astrophysics, high-energy physics, 
etc. A general theory is beautiful, although very abstract \citep{19}. For 
more practical-minded introduction, see  \citep{20}. Naturally many algorithms 
of deconvolution were developed. Our aim in this note was not to suggest one 
more practical algorithm and test its robustness, but indicate that the 
algorithm described in \citep{1} can be naturally generalized for the 
three-dimensional case.

We hope that this refinement of the Ulmer and Kaissl method can find some 
practical applications. Besides many applications in medical radiation physics
\citep{2}, Ulmer and Kaissl method was applied to get a (one-dimensional) 
inverse integral kernel for diffusion in a harmonic potential of an overdamped 
Brownian particle \citep{21}. The method (its three-dimensional version 
described in this article) was also used to obtain a non-local generalization
of the Schr\"{o}dinger equation due to quantum-gravity effects \citep{22}. It
can be also useful in some problems of non-local gravity \citep{23,24}. 

Finally, the relation (\ref{eq43}), which expresses scalar Hermite polynomials 
in terms of ordinary Hermite polynomials can be useful in plasma physics 
(in \citep{9, 10} scalar Hermite polynomials were expressed through the
Laguerre polynomials but the connection with the ordinary Hermite polynomials 
was not noticed).

\section*{Acknowledgments}
We thank Gjergji Zaimi for informing us about \citep{18}. The work of is 
supported by the Ministry of Education and Science of the Russian Federation.

\section*{References}

\bibliographystyle{elsarticle-num}

\end{document}